# Deep Teleportation: Quantum Simulation of Conscious Report in Attentional Blink

Ahmad Sohrabi (ahmadsohrabi@cunet.carleton.ca)
Cognitive Science Department, Carleton University, Ottawa, ON, Canada

Abstract

Recent quantum models of cognition have successfully simulated several interesting effects in human experimental data, from vision to reasoning and recently even consciousness. The latter case, consciousness has been a quite challenging phenomenon to model, and most efforts have been through abstract mathematical quantum methods, mainly focused on conceptual issues. Classical (non-quantum) models of consciousness-related experiments exist, but they generally fail to align well with human data. We developed a straightforward quantum model to simulate conscious reporting of seeing or missing competing stimuli within the famous attentional blink paradigm. In an attentional blink task, a target stimulus (T2) that appears after a previous one (T1) can be consciously reported if the delay between presenting them is short enough (called lag 1), otherwise it can be rendered invisible during the so-called refractory period of attention (lags 2 to 6 and even longer). For modeling this phenomenon, we employed a three-qubit entanglement ansatz circuit in the form of a deep teleportation channel instead of the well-known EPR channel. While reporting the competing stimuli was supposed to be the classical measurement outcomes, the effect of distractor stimuli (i.e., masks, if any) was encoded simply as random angle rotations. The simulation outcome for different states was measured, and the classical outcome probabilities were further used as inputs to a simple linear neural network. The result revealed a non-linear, alternating state pattern that closely mirrors human responses in conscious stimuli reporting. The main result was a successful simulation of Lag 1 sparing, lag 7 divergence, and masking effect through probabilistic outcome of measurement in different conditions.

# 1 Introduction

Our thoughts and attention are changing dynamically instead of remaining fixed and monotonous. These ongoing variations are reflected in brain signals, measured by EEG and other neuroimaging methods. The brain activities are in the form of oscillatory waves observed in the neuroimaging recording of both short and long neural connections, revealing fast and slow frequencies, respectively. The evidence is derived from micro- and macro-level physiological techniques complemented by human performance measures in attentional tasks that investigate various brain activities involved in cognition. These studies have illustrated neurodynamic effects manifested in both periodic and aperiodic cycles and functional connectivities of the brain networks (Gao et al., 2020; Northoff & Huang, 2017).

A widely studied phenomenon that exhibits fluctuations in attention and hence conscious reports of seeing or missing competing stimuli is Attentional Blink (AB). Using typical AB paradigms (e.g., Raymond, Shapiro, & Arnell, 1992; Chun & Potter, 1995), studies have shown that a target stimulus (T2) presented shortly after a previous one (T1) can be consciously reported when the interval is short enough (known as lag 1). Otherwise, it is often rendered invisible during the so-called refractory period of attention (e.g., Nieuwenhuis et al., 2005), spanning lags 2 to 7 (Figure 1). The lag 1 sparing happens if T2 arrives immediately around 100 msec (Gao et al., 2020; Northoff & Huang, 2017). As happens in most studies, Figure 1 also shows a decline from lag 2 to lag 3, followed by an increase beginning at lag 3 onward. However, the result for lag 7 is not consistent across studies as will be described shortly.

Fo be more specific, the detail of a famous study by Nieuwenhuis and colleagues is illustrated in Figure 2, depicting the T2 percent correct results (Nieuwenhuis et al., 2005). Their Figure 6 from two experiments was recreated here to demonstrate when lag 1 sparing occurs, and the blink effect gradually appears from lags 2 to 5 and finally in most conditions it returns to the reportability baseline (no blink) by lags 6 or 7. Another interesting effect can also be found when studies compare the presence or absence of mask stimuli, illustrating a critical divergence at lag 7, with the Mask absent condition showing a reversed pattern compared to Mask present. However, the pattern is not consistent at lag 2 and 3, for example if we compare Figures 1 and 2. Therefore, some inconsistencies exist across studies, especially when attention and masking are involved.

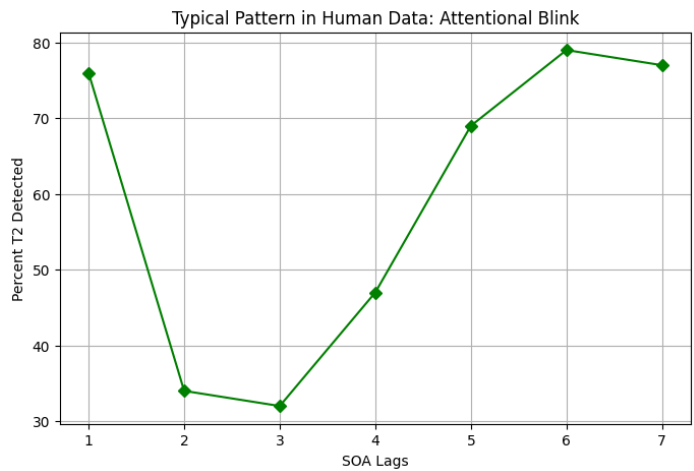

Figure 1. An illustration of a typical pattern in AB based on human data, (based on Nieuwenhuis et al. (2005), their Figure 1A.

Recent quantum models of cognition have focused on several interesting effects found in human experimental data, from vision to reasoning, and recently even consciousness. The latter case, consciousness, has been a quite challenging phenomenon to model. There are a few quantum cognitive models for consciousness related phenomena but they have been using abstract mathematical quantum models, and mainly focused on conceptual issues (e.g., Busemeyer & Lu, 2025). Other previous quantum models have been mostly on verbal, emotional, or social cognition (e.g., Atmanspacher & Filk, 2010; Huang, Cohen, Busemeyer, 2025). Here we aimed to develop a straightforward quantum model of conscious reports of seeing or not seeing competing stimuli as in the famous Attentional Blink (AB) paradigm.

## 1.1 Previous Models

On the modelling side, neurodynamic (classical) models have been used to simulate spikes and oscillations of both single-neuron (Hodgkin & Huxley, 1952) and populations of neurons interactions (Wilson & Cowan, 1972). Based on findings from fast and slow neuromodulatory activities of Locus Coeruleus (LC, e.g., Aston-Jones & Cohen, 2005), neurodynamical models have been able to simulate the AB results including Lag 1 sparing (Nieuwenhuis et al., 2005). According to these studies, the brain typically responds to novel and relevant stimuli in a transient phasic fashion while otherwise showing a tonic and more widespread activation pattern. The phasic response brings a refraction-like period attributed to non-linear activation of neurons and neuromodulatory effects of LC.

The prominent models of neural populations dynamics including LC employ Wilson-Cowan or other dynamic systems that rely on excitation and inhibition (Gilzenrat et al., 2002; Usher & Davelaar, 2002; Sohrabi & West, 2009; see also, Aston-Jones & Cohen, 2005). They are also heavily relying on mathematical formulation of dynamic systems. This makes it an abstract version diverse from other modules which are mainly modelled through spiky or leaky integrated neurons (e.g., Usher & Davelaar, 2002) far from being an integrated model. Also, studies have shown that oscillation can be shown by excitation only especially when refractory period is emphasized (Zhang et al., 2022) so its theoretical and biological basis is not completely known and neurodynamic models made of that might not simulate the underlying mechanisms properly.

Therefore, we aim to provide a quantum model to simulate these effects as a counterpart of classical neurodynamic models. The central question is whether it is possible to construct a quantum model that is more integrated and comprehensive while showing similar behavior to previous models, simulating attentional effects such as fluctuations, oscillations, lag 1 sparing, and lag 7 divergence in Attentional Blink (AB)?

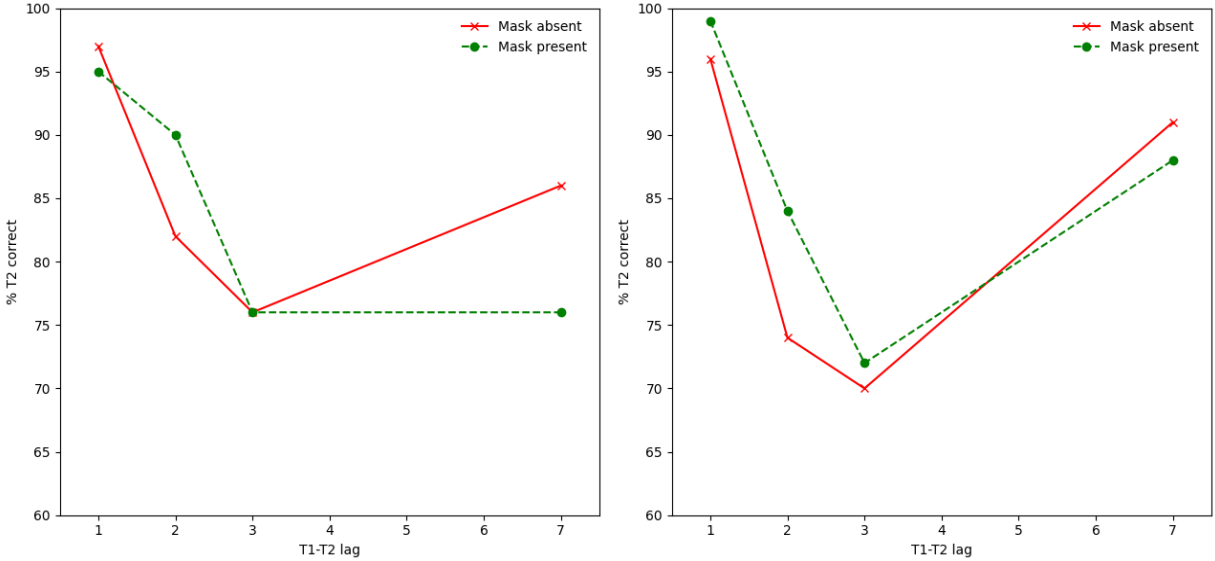

Figure 2. Human data in AB from previous studies. The % T2 correct results of the mentioned study (Nieuwenhuis et al., 2005), their Figure 6, are shown from their Experiment 1 and 2, recreated and depicted here at (Left) and (Right), respectively. It illustrates main effects including lag 1 sparing, critical divergence at lag 7, and masking.

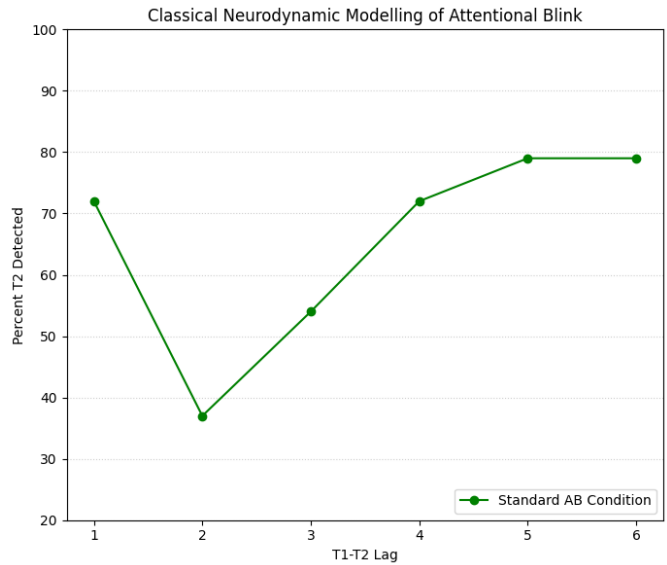

Figure 3. Classical modeling results of AB simulation, recreated from Nieuwenhuis et al. (2005), their Figure 5.

## 1.2 Current Study

For developing a model of the conscious report findings in Attentional Blink (AB), we employed a simple quantum model based on a three-qubit entanglement ansatz in the form of a teleportation circuit. We devised a concept related to what is known as quantum depth (Kaye,

Laflamme, & Mosca, 2007) adopted as sequential depth, as described further in the next section. As will be described further in the next section, the quantum bits were teleported through a novel deep teleportation instead of the well-known single entanglement channel, known as Einstein-Podolsky-Rosen (EPR) channels (Bennett et al., 1993). Two states can be affected by entanglement and making them non separable correlated ($|00>$ and $|11>$) or anti-correlated ($|01>$ and $|10>$) states. While reporting the competing stimuli was supposed to be the classical measurement outcome, the effect of distractor stimuli was encoded simply as random angle rotations.

The simulation outcome for different states was measured, and the classical outcome probabilities were further used as inputs to a simple Feed-Forward Net (FFN) model, for illustration purpose. For the FFN, the bitstrings from the simulations served as inputs while the teleported bit acted as the target. The aim of the whole model was to simulate fluctuations in AB and especially the Lag 1 sparing and masking effects using the probabilistic outcome of the measurement for different conditions. Despite conceptual and theoretical issues, quantum implementation can be an effort toward more integrated models compared to their classical counterpart, including the mentioned neurodynamic model as an example.

## 2 Methods

### 2.1 The model architecture

A popular method in quantum modelling is to encode information in qubits for transformations and operations. The quantum states will be in the form of state vectors consisting of complex numbers. In each ansatz circuit, a series of gates are configured for a specific purpose where superposition and entanglement play the main role. The number of qubits determines the circuit width while the serial arrangement of gates determines its length, made by adding new gates or repeating the same gates. This configuration makes up the final depth of a circuit. Here we call it the entanglement depth of a circuit where its layers involve rotations and control gates. This configuration can be imagined as deep learning layers resembling short and long connectivities. It was also inspired by cognitive tasks, especially the T1-T2 lags as mentioned or similarly Stimulus Onset Asynchrony (SOAs) of stimuli presentation intervals. Therefore, we call it a quantum Cognitive Bio Model (qCBM) which is used along the FFN as called qCBM+FFN (Figure 4).

For our purpose, we apply deep entanglement on the teleportation channel. Three qubits are registered as Qubit (Q), Alice (A), and Bob (B). Then Alice establishes a 'deep entangled' channel on A (as stimuli representation) and B (as stimuli visibility). Alice keeps A and sends B to Bob. She also performs a Bell measurement on Q (as stimuli presentation) and A. Then, based on her measurement outcomes, Bob applies correction based on a designated operation on B and is measured accordingly (as conscious report). For further details, see the next section.

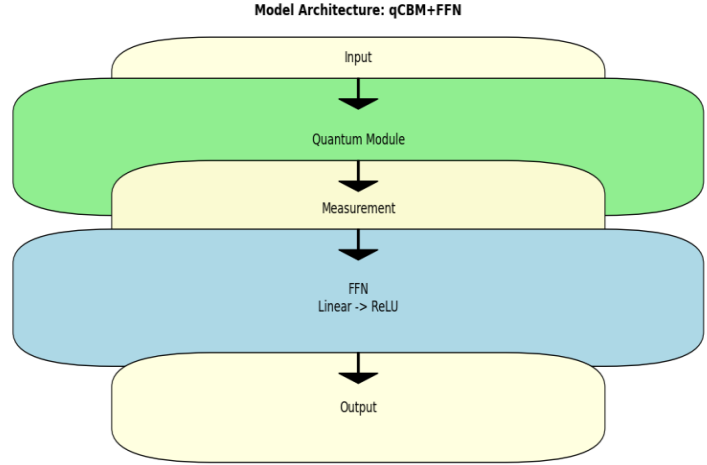

Figure 4. The architecture of the model. The qCBM+FFN consists of quantum and FFN modules. The Measurement component serves as the output of the Quantum module and the input to the FFN.

The FFN component is a simplified version of transformer-like model without attention but only Feed Forward Net (Sohrabi, 2025). This FFN consists of a single linear layer followed by a Rectified Linear Unit (ReLU) with no dropout, implemented using PyTorch (Paszke et al., 2019). It uses simple PyTorch neural net embedding with two vocabulary (0, 1), context length and embed size of 2, and no layer normalization. The popular Adam optimizer is used for gradient-based training with an initial learning rate of 0.001 and a cross-entropy loss function to train and extract probability measures based on a soft max function of logits.

## 2.2 Ansatz details

The teleportation protocol is based on three spin-½ particles/systems representing 3-qubit quantum states ($\Psi$), along three classical registers. For our specific modeling purpose, the teleportation is done through a novel deep teleportation channel instead of the well-known single entanglement channel (Figure 5). As shown in the figure, three qubits are registered as Qubit (Q), Alice (A), and Bob (B), corresponding to the first (already unknown state but with initial exited state using an X gate), second, and third qubit respectively. Then a deep entangled channel is prepared using parametrized CNOT, e.g., starting as $\Phi+$ (Bell state) on A and B. This was implemented through a superposition gate (H) on A followed by a parametrized controlled gate CRX ($\theta$). The rotation was performed with a designated to the second qubit as the target (B). The rotation angles were randomly picked from a fixed interval uniformly between $2\pi/3$ to $4\pi/3$, creating a given rotation range for each run. As noted above, Alice keeps A and sends B to Bob along with the corresponding classical measurement outcomes stored in their respective classical registers. She also performs a Bell measurement on Q and A. Then, based on the outcomes 00, 01, 10, or 11, Bob applies a correction based on the designated operation on B including I (Identity), X, Z, or XZ, respectively and measurement is done accordingly. As Figures 5 shows, the ansatz starts by turning initial state to a starting state 1 (i.e., excited, using X gate) followed by the superposition state (H) on Qubit. Adding a mask was simulated using a U gate with random rotation by Alice ($\theta$ from 0 to $\pi/4$; $\varphi$ and $\lambda$ from 0 to $\pi/2$, see Figure 6). All ansatz circuits were made using IBM Qiskit (2025; see also Javadi-Abhari et al., 2024).

Here by depth, we mean the repeated layers of the entangled teleportation channel, already shown to make deeper encoding as the state vectors are changed gradually (Sohrabi, 2025). For the analysis purpose, we encoded them as Depth Levels ranked from 1 to 14, though they might be different from other depth measures, but almost all measures of depth are linear, as here the Qiskit depth measure of level will be twice the ranks, i.e., 2-28, see also Sohrabi (2025).

At each level of depth, a set of specific gates is repeated to that depth level, from depth 1 to N (1-14) each separated by a barrier just to delineate the series or slices so are just for visualization purposes. Here, the operations do not occur on a physical quantum device, but as an ideal realization is simulated through circuits, having Hermitian and non-Hermitian unitary gates. Hence, the depth resembles the number of layers in variational quantum algorithms (Bravo-Prieto, Lumbreras-Zarapico, Tagliacozzo, & Latorre, 2020).

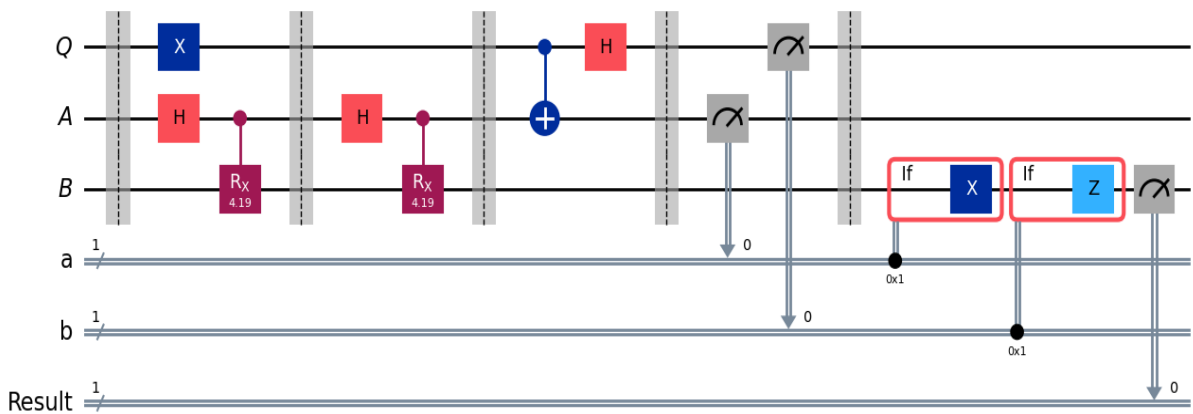

Figure 5. An ansatz circuit used to simulate Attentional Blink (AB) in unmasked condition (Mask absent) showing depth 2 as an example.

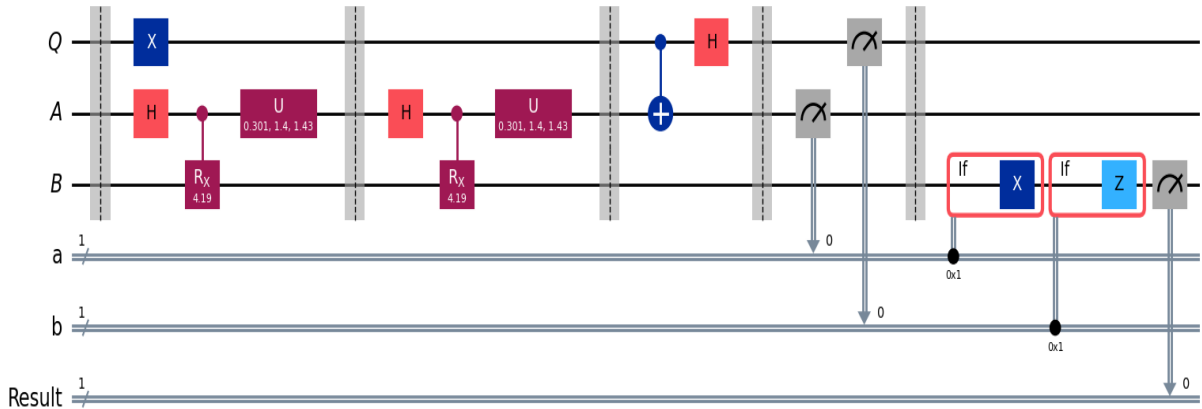

Figure 6. An ansatz circuit used to simulate Attentional Blink (AB) in masked condition (Mask present) with random rotations showing depth 2 as an example.

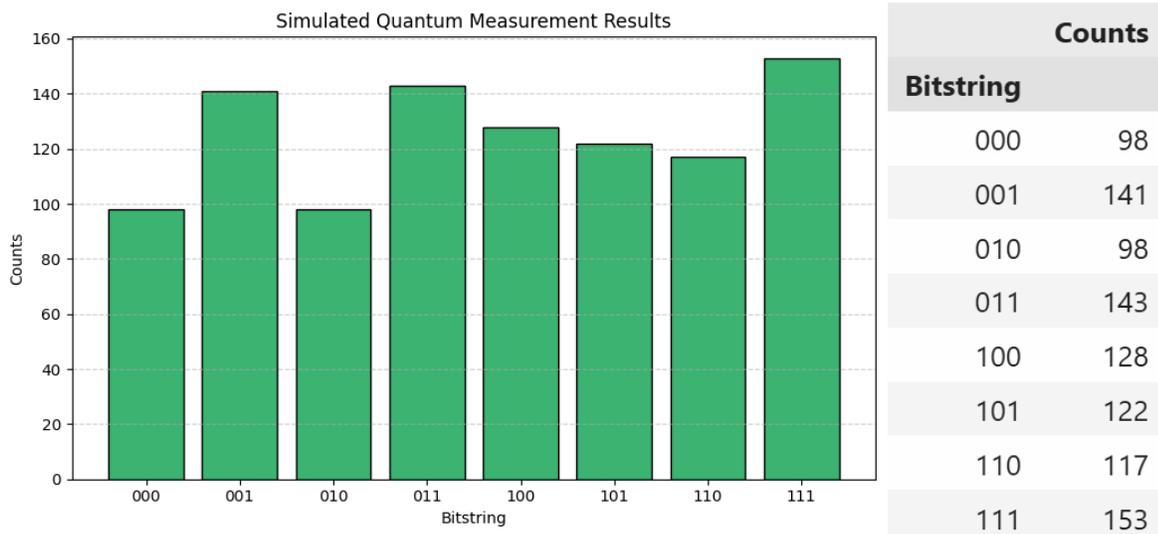

| Bitstring | Counts |
|---|---|
| 000 | 98 |
| 001 | 141 |
| 010 | 98 |
| 011 | 143 |
| 100 | 128 |
| 101 | 122 |
| 110 | 117 |
| 111 | 153 |

Figure 7. An example of the number of counts in 1000 runs at depth 2 for Unmasked Condition (mask absent). The leftmost bitstring digit shows the transported bit used as the target in the later mapping.

3 Results

2.1 Measurement results

The bitstring counts were extracted from Qiskit Aer simulation of the parametrized ansatz circuit at 14 depth levels with or without random qubit rotations as masked or unmasked, respectively. Each condition (depth by masking) was repeated for 1000 runs (to be consistent with typical quantum simulations though small shots work fine) making up an overall 28000 runs. The angles for each rotation were different per simulation but were not meant to introduce noise to the ideal model but to add some variability to mimic experimental human data as reflected in the results of the unmasking versus masking conditions (mask absent vs present). Examples of probability outcomes at depth 2 for the unmasked and masked conditions are shown in Figure 7 and 8, respectively. The leftmost bitstring digit shows the transported bit used as the target in the later mapping while the other two served as inputs.

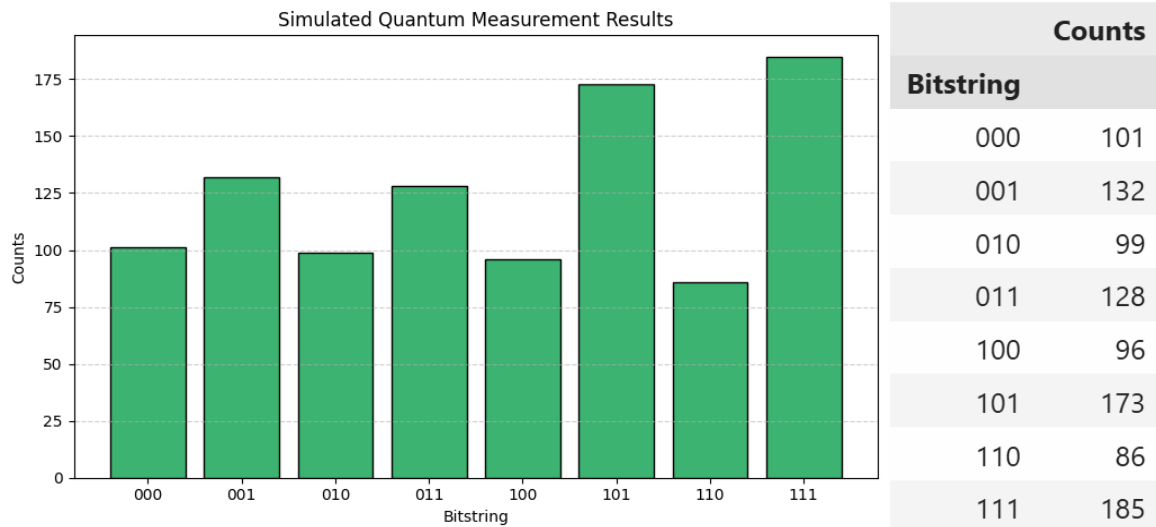

Figure 8. An example of the number of counts in 1000 runs at depth 2 for Masked Condition (mask present). The leftmost bitstring digit shows the transported bit used as the target in the later mapping.

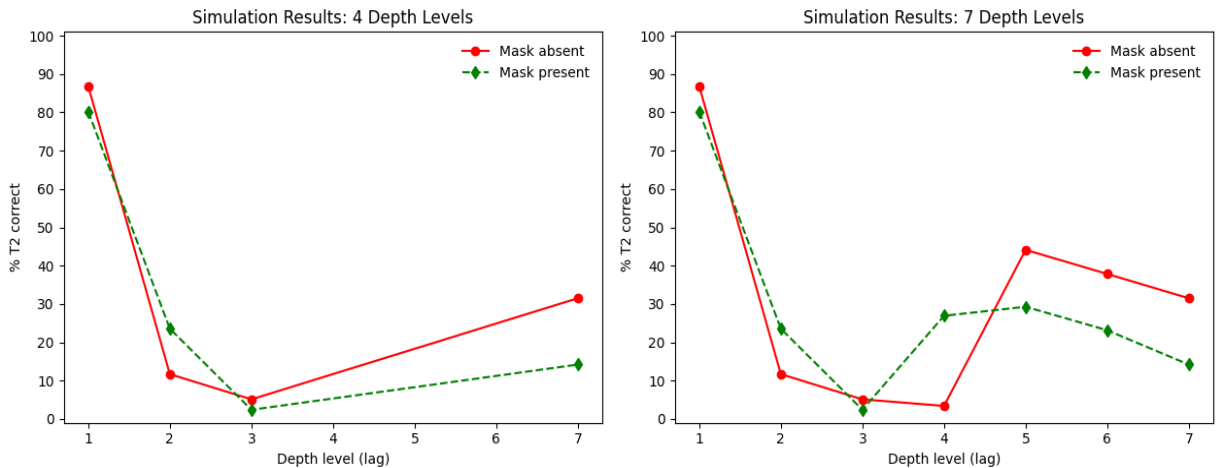

Figure 9. Main simulation results. (Left) Like human data at 4 lags (Figure 1-2), the results show that lag 1 is spared and lag 7 recovers, especially in Mask absent condition. (Right) Same results for all 7 lags. The overall pattern is similar to the well-known u-shape trend in human studies (see Figure 1), compared with Mask present condition (as with human data in Figure 2).

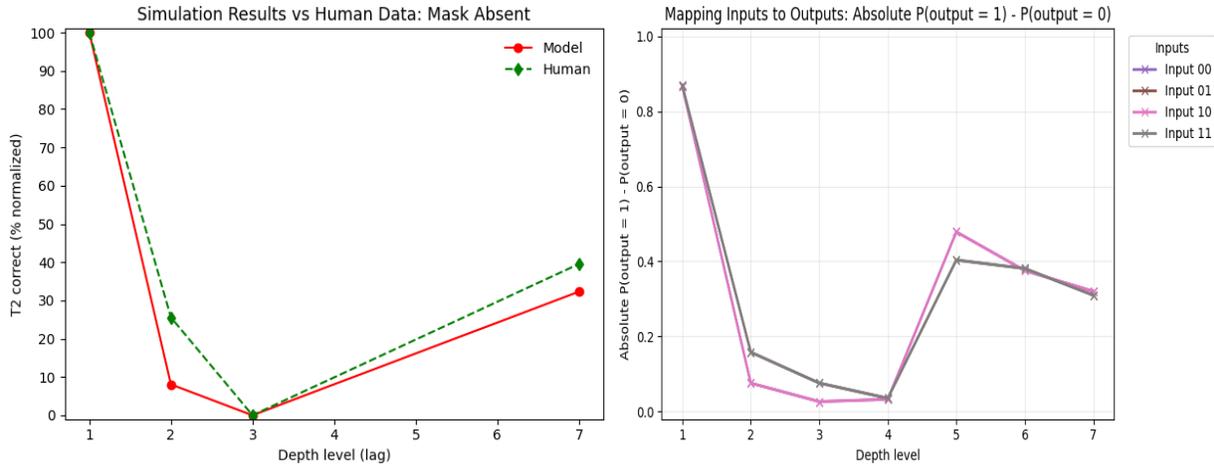

Figure 10. Simulation results of the unmasked condition (Mask absent). (Left) Simulation results compared to human data mentioned in Figure 2, at depths 1-3 and 7 as lags for seen (p (output = 1)) minus unseen (p (output = 0), absolute value. (Right) Same results, Simulation only from which the Left figure was created by averaging across states. Here bitsrings for all depths/lags up to 7 are shown, but only two states can be seen, overlapped by the other two.

## 3.1 Neural Net Mapping Results

We calculated the probability of output for each state based on the FFN output, after 300 epochs, though not much learning was needed for the current stable mapping. It gave probabilities of 1, i.e., P(output=1) as T2 seen versus 0, i.e., P(output=0) as T2 unseen for both correlated (00, 11) and anticorrelated (01, 10) states. We provide all results but first we focus on the main effects found in human data mentioned above, by measuring T2 report percentage by subtracting P(output=0) from P(output=1) and taking its absolute values across conditions (the real numbers we multiplied by 100 to have % correct scores).

As shown in Figure 9, the modelling results are similar to what was mentioned previously about human studies on Attentional Blink (AB). Specifically, it simulated the results from Nieuwenhuis et al. (2005), Experiment 1, their Figure 6 (see Figure 2). It depicts the lag 1 sparing and a critical divergence at lag 7 in the Mask absent compared to Mask present condition showing a reversed pattern. In the human data (Figures 1-2), lag 1 is spared and lag 7 recovers (spared), especially in Mask absent condition while the reportability at other lags is changing too. As mentioned before, the pattern is not consistent at lag 2 and 3, for example in the results of studies mentioned above and shown in Figures 1 and 2. Therefore, that inconsistency across studies was reflected in the current simulation results as well, to be elaborated further in the discussion. As will be described in the next section, the details are shown in Figures 10-14 for unmasked condition (mask absent) and in Figures 15-19 for masked condition (mask present).

## 3.2 Details of Masked and Unmasked Results

As shown in Figure 10, the model could simulate the data in a human study previously mentioned in Figure 2. It shows similarities at depths 1-3 and 7 as lags measured as seen, i.e., p (output = 1) minus unseen i.e., p (output = 0). It also depicts the same simulation results averaged across all separately without combining the states. The same results for all depths/lags up to 7 are also shown. We can see that the overall pattern is similar to the well-known u-shape trend in human studies (see Figure 1). The parameters (e.g., angle rotations) can be adjusted to achieve a closer fit with human data.

Simulation results from which Figure 9 were extracted are shown in Figure 11 again as lags but for both seen, p (output = 1), and unseen, p (output = 0), separately not their subtraction, showing all the first 7 depths. The results of all 14 depth levels are illustrated in Figure 12. Overall, they show the probabilities of seen vs unseen that are reversed around depth/lag 3 and the difference oscillates and vanishes by deeper lags. We further analyzed the data by combining the same data from Figure 12 for consecutive pairwise depths merged to make seven depths instead of 14 (Figure 13). Again, the overall pattern was similar to the well-known u-shape trend in human studies (see Figure 1). Finally, Figure 14 shows the results of the unmasked condition illustrated as graph networks for nine depths, showing the probabilities of seen, p (output = 1), and unseen, p (output = 0) that are reversed around depth/lag 3 and the difference vanishes by deeper lags. The rest of the figures illustrate the same results now for the masked condition (Figures 15-19). As noted above, the main differences include more oscillations, especially at deeper levels, reflecting the lag 7 divergence of masking effect like human data from Figure 2.

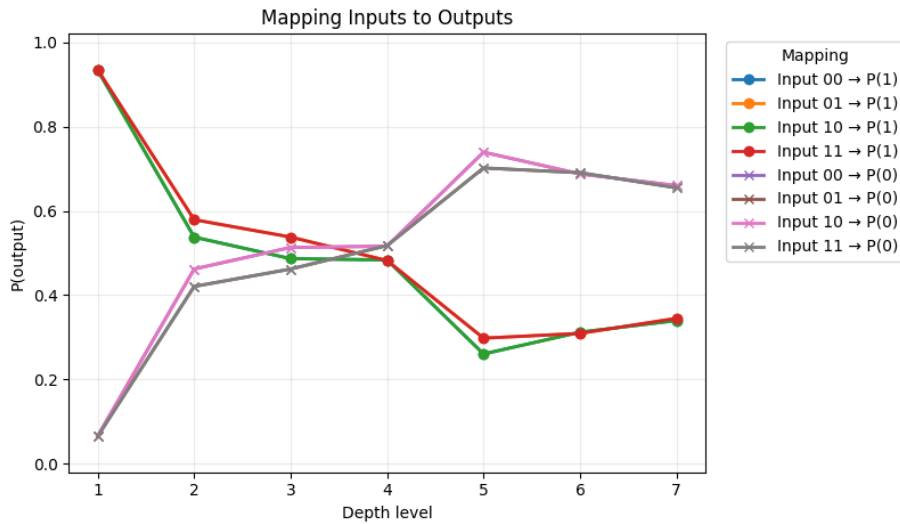

Figure 11. Simulation results from which Figure 9 were extracted again as lags but for both seen (p (output = 1)) and unseen (p (output = 0), separately not their subtraction. So here all first 7 depths are shown. Note: only one of the two corelated is shown as they are overlapped, same as anticorrelated states.

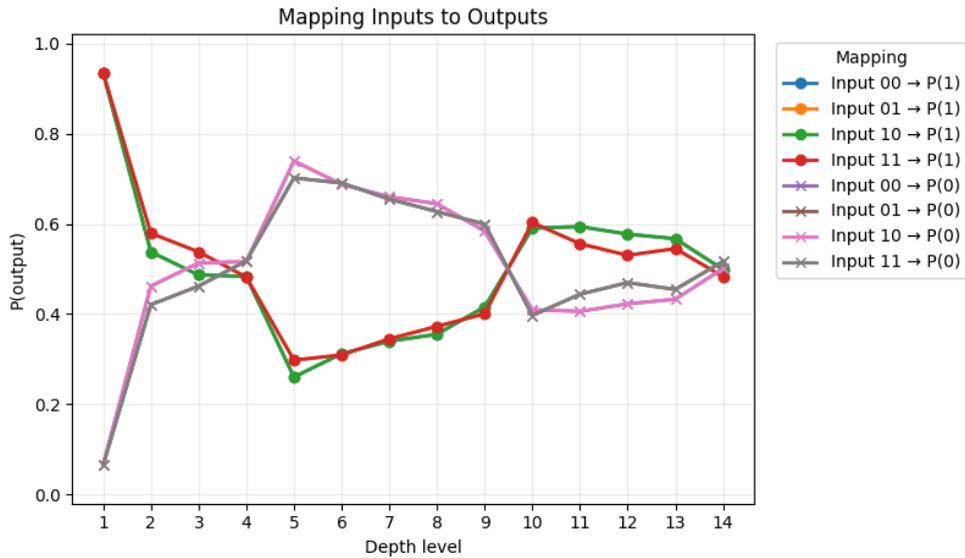

Figure 12. Simulation results from which Figure 10-11 were extracted again as lags but for both seen (p (output = 1)) and unseen (p (output = 0), separately not their subtraction. But here all 14 depths are shown. Overall, it shows the probabilities of seen vs unseen that are reversed around depth/lag 3 and the difference oscillates and vanishes by deeper lags. Note: only one of the two corelated is shown as they are overlapped, same as anticorrelated states.

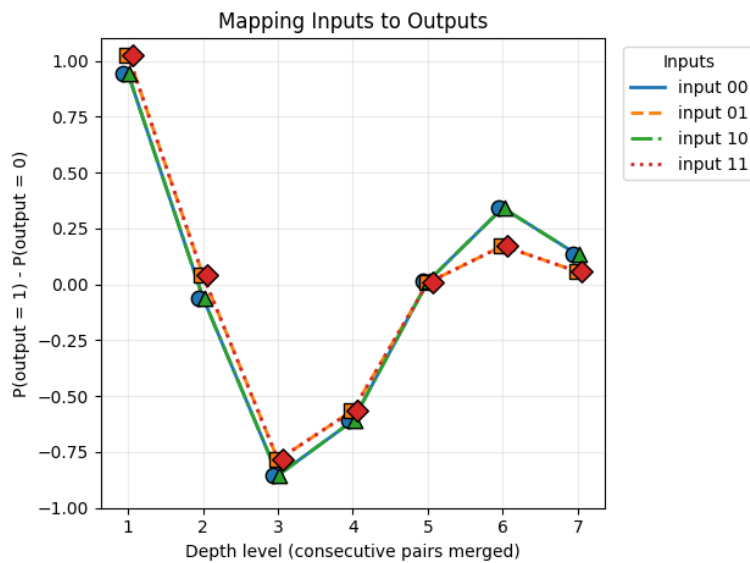

Figure 13. Results of the unmasked condition simulation depicting the same data as Figure 17 where consecutive pairwise depths were merged to make seven depths instead of 14. All correlated and anticorrelated states are shown but as subtraction of seen, p (output = 1), and unseen, p (output = 0), not separately. The overall pattern is similar to the well-known u-shape trend in human studies (see Figure 1).

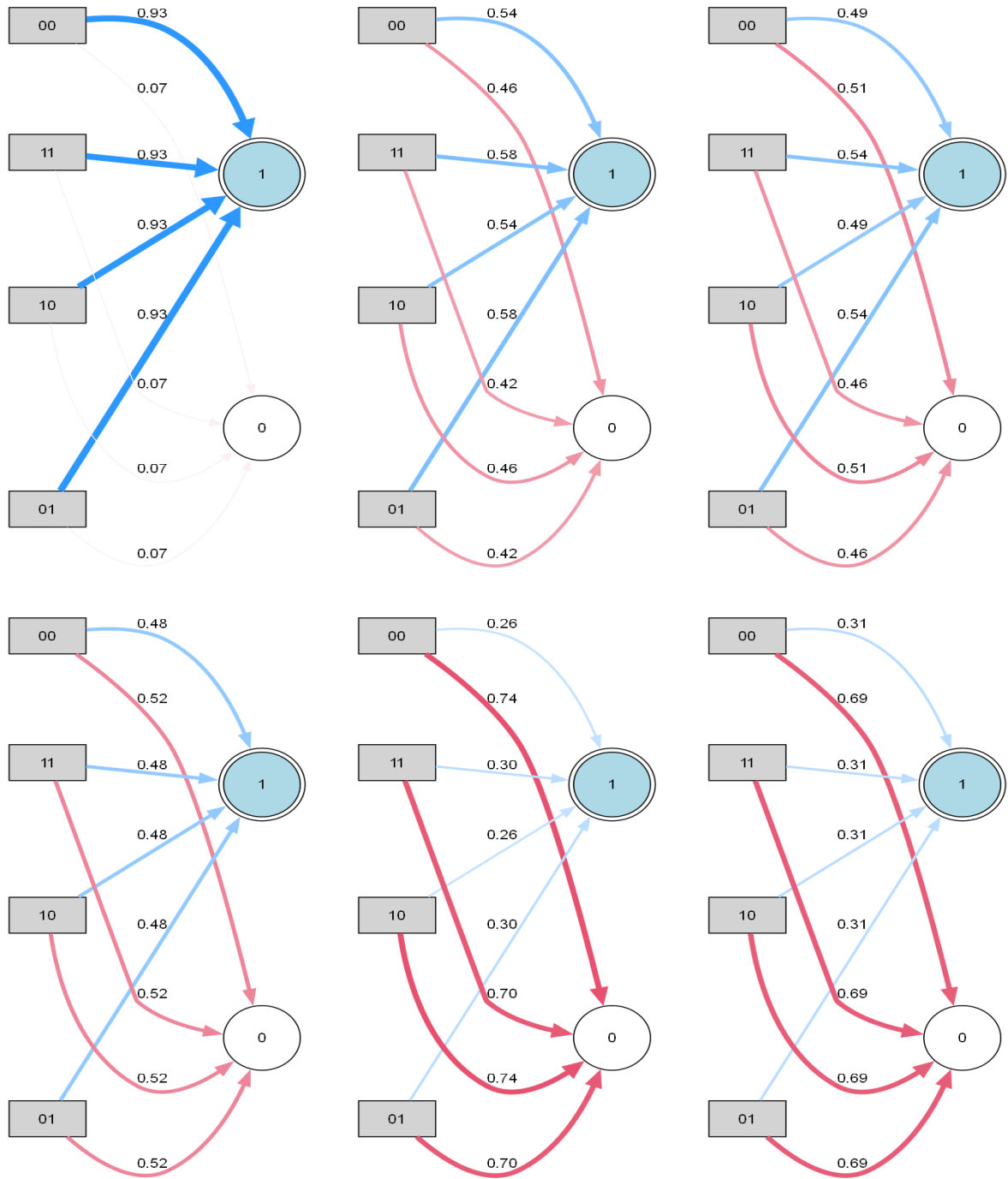

Figure 14. Results of the unmasked condition illustrated as graph networks for nine depths, showing the probabilities of seen (1) vs unseen (0) that are reversed around depth/lag 3 and the difference vanishes by deeper lags (see Figure 12).

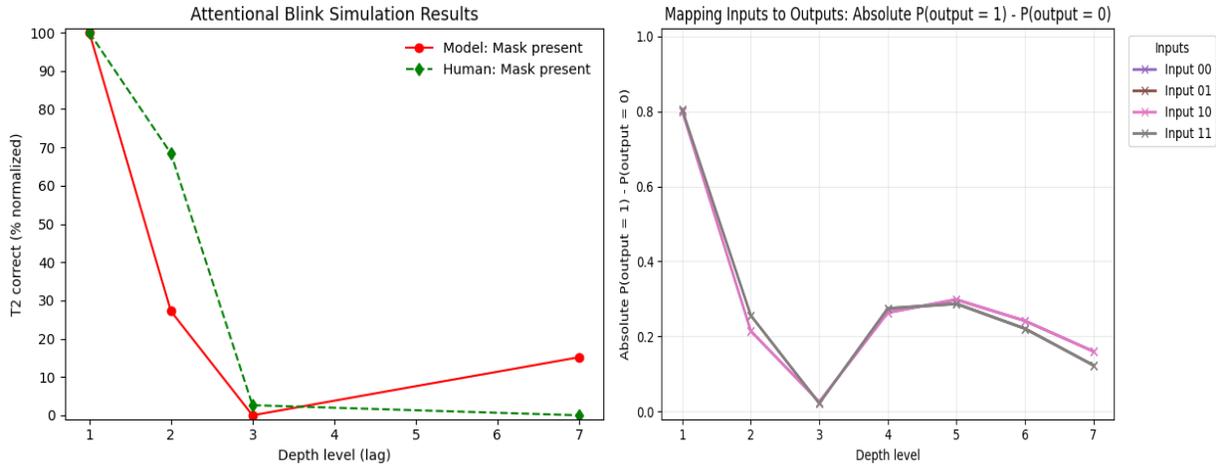

Figure 15. Simulation results of the masked condition (Mask present). (Left) Simulation results compared to human data mentioned in Figure 2, at depths 1-3 and 7 as lags for seen (p (output = 1)) minus unseen (p (output = 0)). (Right), absolute value. Same results, Simulation only from which the Left figure was created by averaging across states. Here bitsrings for all depths/lags up to 7 are shown, but only two states can be seen, overlapped by the other two.

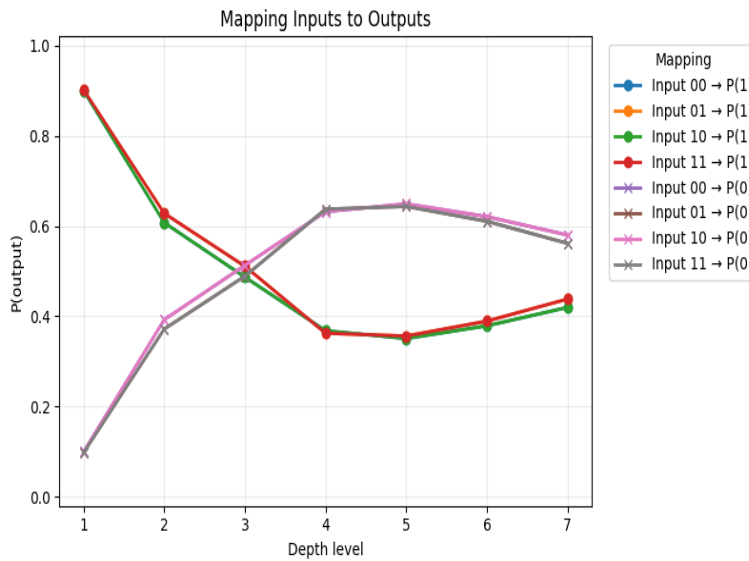

Figure 16. Results of the masked condition simulation, from which Figure 15 were extracted again as lags but for both seen (p (output = 1)) and unseen (p (output = 0), separately not their subtraction. So here all the first 7 depths are shown. Note: only one of the two corelated is shown as they are overlapped, same as anticorrelated states.

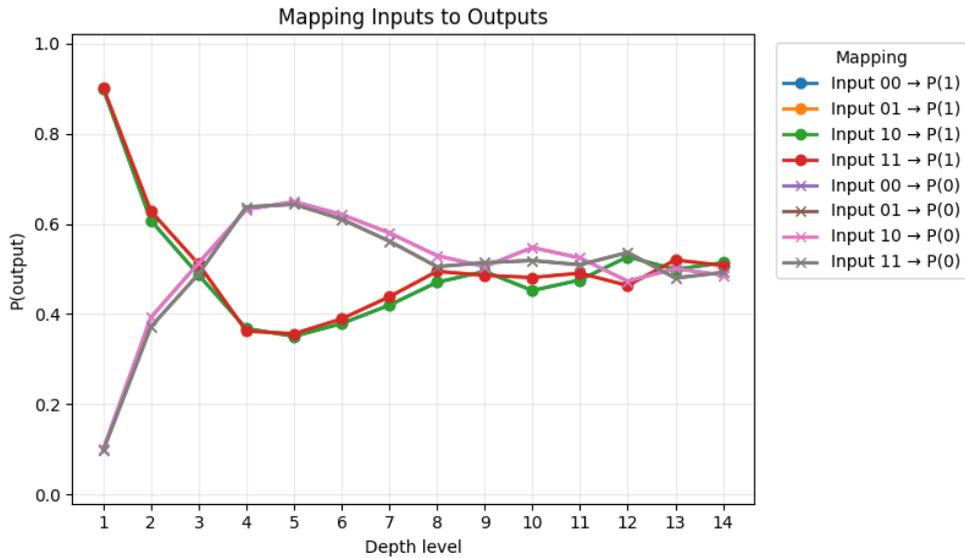

Figure 17. Results of the masked condition simulation, from which Figure 15-16 were extracted again as lags but for both seen (p (output = 1)) and unseen (p (output = 0), separately not their subtraction. But here all 14 depths are shown. Overall, it shows the probabilities of seen vs unseen that are reversed around depth/lag 3 and the difference oscillates and vanishes by deeper lags. Note: only one of the two corelated is shown as they are overlapped, same as anticorrelated states.

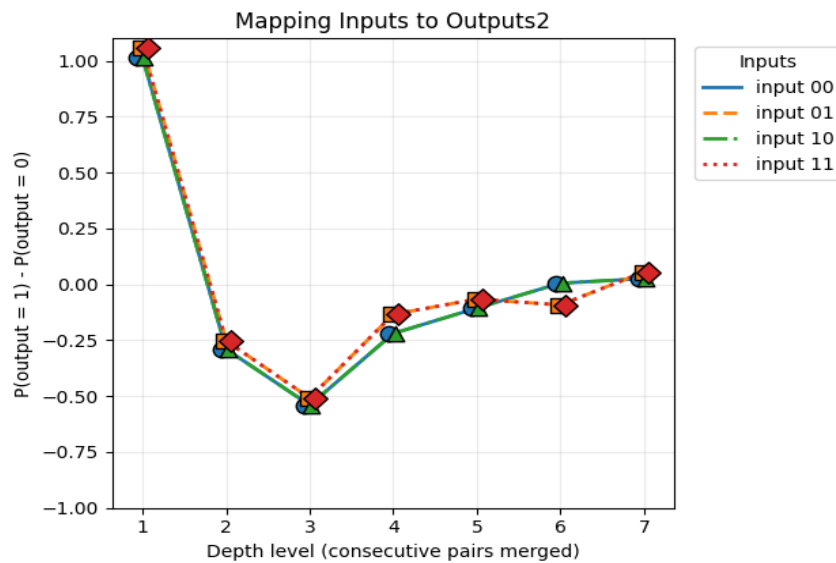

Figure 18. Results of the masked condition simulation depicting the same data as Figure 17 where consecutive pairwise depths were merged to make seven depths instead of 14. All correlated and anticorrelated states are shown but as a subtraction of seen, p (output = 1), and unseen, p (output = 0), not separately. The overall pattern is similar to the well-known u-shape trend in human studies, but with a long right tail (see Figure 2).

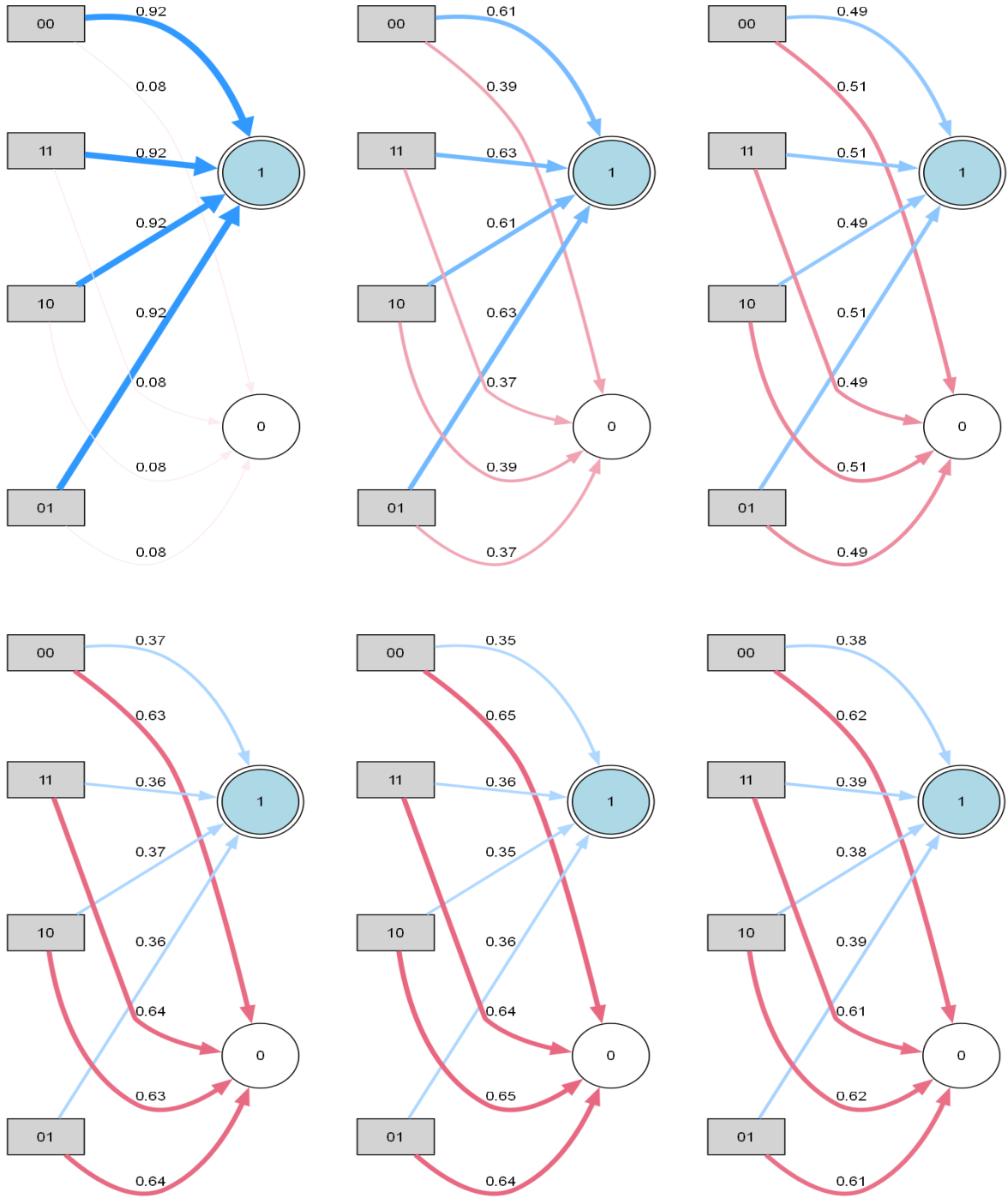

Figure 19. Results of masked condition simulation, illustrated as neural net-like graphs for six depths. Overall, it shows the probabilities of seen (1) vs unseen (0) that are reversed around depth/lag 3 and the difference oscillates and vanishes by deeper lags (see Figure 17).

## 4 Discussion

The findings of the current study revealed applicability of quantum-based cognitive models of consciousness relevant experiments apparently better than its classical neurodynamic models. We employed a simulation of the well-known three-qubit entanglement circuit but through a deep teleportation instead of a single entanglement channel, known as Einstein-Podolsky-Rosen (EPR) channels (Bennett et al., 1993). This method provided state trajectories similar to neurodynamic events in the brain, studied using cognitive tasks and neurophysiology. While the abovementioned dynamic models consist of both spiky neurons and abstract mathematical dynamics, the current model was based on straightforward and integrated methods in quantum simulations.

As in the human studies on attentional blink, the model showed fluctuations related to effects such as lag 1 sparing and lag 7 divergence, as well as masking through probabilistic outcome of measurement in different conditions. Another effect overlooked by classic neurodynamical models (Nieuwenhuis et al., 2005; Sohrabi & West, 2009) is the change between lag 2 and 3. As we can see by comparing human data (Figure 2-3) to previous modeling results (Figure 3), the previous models do not show a drop from lag 2 to 3 but a sharp rise starting from lag 2. This effect was present in the current model result along with other main effects, as for example in Figure 9, similar to human data (Raymond, Shapiro, & Arnell, 1992; Chun & Potter 1995; Nieuwenhuis et al., 2005).

Although most studies attribute these fluctuations and inconsistencies to cortical activities, deeper brain layers can have effects as well. These effects have been mainly attributed to a limitation of our attentional resources, mainly in the Locus Coeruleus (LC). This attentional refractory period or bottleneck is similar to that in single neuron refractory period and in populations of neurons modeled by classical neurodynamic models of AB (Nieuwenhuis et al., 2005) and other cognitive tasks such as masked priming (Sohrabi & West, 2009) and Stroop (Sohrabi & West, 2025) to name but a few. However, the attentional and other fluctuations are not restricted to LC and other brain areas are also involved including hypothalamus (Nair et al., 2025). So, a more comprehensive picture of the underlying neural mechanisms requires further investigation.

In the current model we can see the relation of conscious report to the wave-like oscillations, i.e., amplitudes measured by probabilities, similar but, as mentioned, better than previous classical neurodynamic models (Nieuwenhuis et al., 2005; Sohrabi & West, 2009). So, we are taking these results as a simulation of conscious reports, not conscious experience which might not necessarily result from the "collapse of wave function" (Hameroff and Penrose, 1996). Other quantum models have been implemented using abstract mathematical methods and are mainly devoted to high level human data such as verbal, emotional, or social cognition (Atmanspacher & Filk, 2010; Huang, Cohen, Busemeyer, 2025; Busemeyer & Lu, 2025).

As mentioned, the subcortical brain area Locus Coeruleus (LC) has long been known to be involved in attention-related phenomena. Previous neurodynamic models are mainly based on neurophysiological data related to LC (e.g., Gilzenrat et al., 2002; Usher & Davelaar, 2002). Although, recent arguments suggest that consciousness may in fact be grounded in subcortical activities of the so-called reptilian brain (Melloni et al., 2021) but cognitive experimental studies confirm the interactive effects of higher cognitions adding to attentional phenomena and making the results unstable and less predictable. For example, AB is affected by other processes such as memory load (Akyürek & Hommel, 2005), which is involved in cortical activities (e.g., Goldman-Rakic, 1995; Degutis et al., 2024). Therefore, achieving a comprehensive understanding will require further interdisciplinary investigations.

The quantum depth utilized in this study warrants further investigation to determine whether it emerges simply from neuronal layers or other biological compartmental configurations. For example, in their "Orch OR" (Orchestrated Objective Reduction) model, Hameroff and Penrose (1996) argue that consciousness arises from "orchestrated" collapse of wavefunction in arrays of microtubule crystals in the neurons. Instead, the current model shows the behaviours found in detailed cognitive processes involving precise measures of presentation timing, reaction times, and errors. Although we showed results for both 7 lags and 14 lags, reflecting scale invariance u-shaped patterns found in previous human and modeling studies, we know that studies and brain dynamics vary in terms of experimental set up and task instructions. This adds to repetition issue in psychological experiment although it is much less in cognitive experiments, but they have their own issues including inconsistency or fragility between studies (Sohrabi & West, 2025) or even between experiments in a single study as mentioned about (Nieuwenhuis et al., 2005). This makes a challenge for a better fit between human and model data, especially ideal quantum simulations, so additional research is needed for a more detailed comparison.

## Acknowledgements

The author acknowledges comments provided by Andrew Brook on conceptual issues related to consciousness.

## References

Akyürek, E., & Hommel, B. (2005). Direct evidence for a role of working memory in the attentional blink. *Memory & Cognition, 33*(4), 597–604. https://doi.org/10.3758/BF03195329

Atmanspacher, H., & Filk, T. (2010). A proposed test of temporal nonlocality in bistable perception. *Journal of Mathematical Psychology, 54*(3), 314–321. https://doi.org/10.1016/j.jmp.2009.12.001

Aston-Jones, G., & Cohen, J. D. (2005). An integrative theory of locus coeruleus-norepinephrine function: Adaptive gain and optimal performance. *Annual Review of Neuroscience, 28*, 403–450. https://doi.org/10.1146/annurev.neuro.28.061604.135709

Bennett, C. H., Brassard, G., Crépeau, C., Jozsa, R., Peres, A., & Wootters, W. K. (1993). Teleporting an unknown quantum state via dual classical and Einstein-Podolsky-Rosen channels. *Physical Review Letters, 70*(13), 1895–1899. https://doi.org/10.1103/PhysRevLett.70.1895

Bravo-Prieto, C., Lumbreras-Zarapico, J., Tagliacozzo, L., & Latorre, J. I. (2020). Scaling of variational quantum circuit depth for condensed matter systems. *arXiv preprint arXiv:2002.06210*. https://doi.org/10.48550/arXiv.2002.06210

Busemeyer, J. R., & Lu, M. (2025). Quantum consciousness, brains, and cognition. *Journal of Consciousness Studies, 32*(9–10), 156–182. https://doi.org/10.53765/20512201.32.9.15

Degutis, J. K., Chaimow, D., Haenelt, D., Assem, M., Duncan, J., Haynes, J.-D., Weiskopf, N., & Lorenz, R. (2024). Dynamic layer-specific processing in the prefrontal cortex during working memory. *Communications Biology, 7*, Article 6780. https://doi.org/10.1038/s42003-024-06780-8

Gao, R., van den Brink, R. L., Pfeffer, T., & Voytek, B. (2020). Dynamic topographies of intrinsic neural timescales: A key role for consciousness. *Journal of Mathematical Psychology, 95*, 102327. https://doi.org/10.1016/j.jmp.2020.102327

Gilzenrat, M. S., Holmes, B. D., Rajkowski, J., Aston-Jones, G., & Cohen, J. D. (2002). Simplified dynamics in a model of noradrenergic modulation of cognitive performance. *Neural Networks, 15*(4–6), 647–663. https://doi.org/10.1016/S0893-6080(02)00055-2

Goldman-Rakic, P. S. (1995). Cellular basis of working memory. *Neuron, 14*(3), 477–485. https://doi.org/10.1016/0896-6273(95)90304-6

Hameroff, S., & Penrose, R. (1996). Orchestrated reduction of quantum coherence in brain microtubules: A model for consciousness. *Philosophical Transactions of the Royal Society A, 354*(1711), 1929–1936. https://doi.org/10.1098/rsta.1996.0056

Hodgkin, A. L., & Huxley, A. F. (1952). A quantitative description of membrane current and its application to conduction and excitation in nerve. *The Journal of Physiology, 117*(4), 500–544. https://doi.org/10.1113/jphysiol.1952.sp004764

Huang, J., Cohen, J., & Busemeyer, J. (2025). A quantum model of arousal and Yerkes-Dodson law. In Proceedings of the Annual Meeting of the Cognitive Science Society (Vol. 47). Cognitive Science Society.

IBM Quantum. (2023). IBM Quantum Cloud Platform. https://quantum.cloud.ibm.com

Javadi-Abhari, A., Treinish, M., Krsulich, K., Wood, C. J., Lishman, J., Gacon, J., et al. (2024). Quantum computing with Qiskit. *arXiv preprint arXiv:2405.08810*. https://doi.org/10.48550/arXiv.2405.08810

Kaye, P., Laflamme, R., & Mosca, M. (2007). Quantum algorithms and applications. Batista Lab, Yale University. https://files.batistalab.com/teaching/attachments/chem584/Mosca.pdf

Melloni, L., Mudrik, L., Pitts, M., & Koch, C. (2021). Making the hard problem of consciousness easier. *Science, 372*(6545), 911–912. https://doi.org/10.1016/j.tics.2021.03.002

Nair, A., Vinograd, A., Liu, M., Mountoufaris, G., Linderman, S., & Anderson, D. J. (2025, December 3). The neural computation of affective internal states in the hypothalamus: A dynamical systems perspective. *Neuron, 113*(23), 3887–3907. https://doi.org/10.1016/j.neuron.2025.11.003

Nieuwenhuis, S., Gilzenrat, M. S., Holmes, B. D., & Cohen, J. D. (2005). The role of the locus coeruleus in the regulation of cognitive performance. *Journal of Experimental Psychology: General, 134*(3), 291–307. https://doi.org/10.1037/0096-3445.134.3.291

Northoff, G., & Huang, Z. (2017). Connecting brain and mind through temporo-spatial dynamics: Towards a theory of common currency. *Journal of Mathematical Psychology, 78*, 20–29. https://doi.org/10.1016/j.jmp.2017.06.004

Paszke, A., Gross, S., Massa, F., et al. (2019). PyTorch: An imperative style, high-performance deep learning library. *Advances in Neural Information Processing Systems*.

Sohrabi, A. (2025). Quantum depth for evaluating transformer-like models. *Carleton University Cognitive Science Technical Report 2025-02*.

Sohrabi, A., & West, R. L. (2009). Positive and negative congruency effects in masked priming: A neuro-computational model based on representation, attention, and conflict. *Brain Research, 1289*, 124–132. https://doi.org/10.1016/j.brainres.2009.06.091

Sohrabi, A., & West, R. L. (2025, March 24). The fragility of the reverse facilitation effect in the Stroop task: A dynamic neurocognitive model. *arXiv preprint arXiv:2503.19128*. https://arxiv.org/abs/2503.19128

Usher, M., & Davelaar, E. J. (2002). Neuromodulation of decision and response selection. *Neural Networks, 15*(4–6), 635–645. https://doi.org/10.1016/S0893-6080(02)00054-0

Wilson, H. R., & Cowan, J. D. (1972). Excitatory and inhibitory interactions in localized populations of model neurons. *Biophysical Journal, 12*(1), 1–24. https://doi.org/10.1016/S0006-3495(72)86068-5

Zhang, Y., Li, X., Wang, Z., & Chen, H. (2022). Oscillatory dynamics in excitatory neural populations with refractory periods. *arXiv preprint arXiv:2204.00583*. https://doi.org/10.48550/arXiv.2204.00583